\documentstyle[prb,aps]{revtex}

\begin{document}

\draft

\title{Microscopic Analysis of the Non-Dissipative Force on a Line
        Vortex in a Superconductor}
\author{Frank Gaitan}
\address{International Center for Theoretical Physics\\
         P. O. Box 586\\
         34100 Trieste\\
         ITALY}
\date{\today}
\maketitle

\begin{abstract}
A microscopic analysis of the non-dissipative force ${\bf F}_{nd}$
acting on a line vortex in a type-II superconductor at $T=0$ is
given. All work presented assumes a charged BCS superconductor.
We first examine the Berry phase induced in the BCS superconducting
ground state by movement of the vortex and show how this phase
enters into the hydrodynamic action $S_{hyd}$ of the
superconducting condensate. Appropriate variation of $S_{hyd}$
gives ${\bf F}_{nd}$ and variation of the Berry phase term is seen
to contribute the Magnus or lift force of classical hydrodynamics
to ${\bf F}_{nd}$. This analysis, based on the BCS ground state of
a {\em charged\/} superconductor, confirms in detail the arguments of
Ao and Thouless within the context of the BCS model. Our Berry
phase, in the limit $e\rightarrow 0$, is seen to reproduce the
Berry phase determined by these authors for a {\em neutral\/} superfluid.
We also provide a second, {\em independent\/}, determination of
${\bf F}_{nd}$ through a microscopic derivation of the continuity
equation for the condensate linear momentum. This equation yields
the acceleration equation for the superflow and shows that the vortex
acts as a sink for the condensate linear momentum. The rate at which
momentum is lost to the vortex determines ${\bf F}_{nd}$ in this
second approach and the result obtained agrees identically with the
previous Berry phase calculation. The Magnus force contribution to
${\bf F}_{nd}$ is seen in both calculations to be a consequence of
the vortex topology and motion.
\end{abstract}

Already in the phenomenological/macroscopic models of
vortex dynamics in type-II superconductors due to Bardeen-Stephen (BS)
and Nozi\`{e}res-Vinen (NV) \cite{phen}, the form of the
non-dissipative force ${\bf F}_{nd}$ acting on the vortex is
controversial. This force is the result of the vortex's interaction with
an applied magnetic field ${\bf H}_{ext}$, an electric field
${\bf E}$ due to the vortex motion, and the
surrounding condensate of superconducting electrons. The disagreement
centers on whether the vortex feels the lift or Magnus force of classical
hydrodynamics as a consequence of its motion through the superconducting
condensate. In the BS model, the non-dissipative force is due strictly
to the Lorentz force
$\rho_{s} h \omega ({\bf v}_{s}\times\hat{{\bf z}})/2$;
while in the NV model, the Lorentz force is supplemented by the Magnus
force $-\rho_{s}m{\cal K}{\bf v}_{s}\times\hat{{\bf z}}$ \cite{defs}.
In a very interesting paper, Ao and Thouless \cite{ath} have returned
to this controversy arguing that the correct form for ${\bf F}_{nd}$
is the NV-form, and that the Magnus force contribution to it is a
manifestation of a Berry phase induced in the many-body ground state
due to the vortex motion. They provide a calculation for a neutral
superfluid and argue that the same scenario will also apply for a charged
superconductor. Given that the BCS model of superconductivity provides a
highly successful microscopic description of the dynamics of a
{\em charged\/} superconductor, it would be very interesting to see if
${\bf F}_{nd}$ can be determined using this model of a charged superconductor
(together with the
starting assumptions common to BS and NV, see below). In this Brief Report
we report the results of two such calculations. A detailed presentation
and discussion of these calculations will be reported elsewhere \cite{me}.
In the first calculation we determine ${\bf F}_{nd}$ by working with
the BCS superconducting ground state in the case where a vortex is
present. This state is first constructed and the Berry phase induced in it
by the vortex motion is determined. We then show how this Berry phase
enters into the action describing the hydrodynamic degrees of
freedom of the superconducting condensate. Variation of this
action with respect to the vortex trajectory gives ${\bf F}_{nd}$ and
the result found is seen to take the NV-form. In the second calculation
we give a microscopic derivation of the acceleration equation for the
superflow. Together with the expected contributions off the vortex due to
spatial variation of the chemical potential, and the electric and magnetic
fields present, we also find a singular term arising from the vortex
topology which describes the disappearance of linear momentum into the
vortex. The rate at which this momentum is disappearing gives
${\bf F}_{nd}$ and is found to agree identically with the result of the
Berry phase calculation. We stress that the two calculations are
{\em independent\/}
of each other, and each shows that the Magnus force contribution
to ${\bf F}_{nd}$ arises as a consequence of the vortex topology.

We make use of the Bogoliubov equation to treat the superconducting
dynamics. The gap function takes the form $\Delta ({\bf r})=\Delta_{0}(r)
\exp [-i\theta ]$ in the presence of a line vortex with winding number
$\omega = -1$ (in cylindrical coordinates ($r$, $\theta$, $z$) centered
on the vortex). As in the models of BS and NV, we: (i) assume $T=0$; (ii)
will approximate the non-local character of BCS superconductivity by a
local dynamics; (iii) assume $H_{c_{1}}<H_{ext}\ll H_{c_{2}}$ so that
vortex-vortex interactions can be ignored and attention can focus on a
single vortex; (iv) assume a clean type-II superconductor so that pinning
effects can be ignored; and, (v) set $\hbar=m=c=1$ unless
otherwise stated. The solutions of the Bogoliubov equation in the presence
of a line vortex are well-know \cite{car} and can have positive and
negative energies relative to the Fermi energy. The superconducting
ground state is constructed by occuppying the negative energy states. The
charge conjugation degree of freedom for the two-component Nambu
quasi-particle (NQP) is labeled by $2s_{z}$, and the operator that
{\em creates} a negative energy NQP is $\gamma_{n\downarrow}$ (where $n$
labels the energy spectrum). Thus, the ground state in the presence of a
vortex is
\begin{equation}
|BCS\,\rangle = \prod_{n} \gamma_{n\downarrow}|0\,\rangle
  \label{grd}\hspace{0.1in}.
\end{equation}
$\gamma_{n\downarrow}$ depends linearly on the (complex conjugate) of the
components of the solutions of the Bogoliubov equation ($u_{n}$, $v_{n}$)
\cite{car}. Adiabatic motion of the vortex generates a Berry phase \cite{ber}
$\phi_{n}$ in the solutions ($u_{n}$, $v_{n}$). Consequently, $\gamma_{n
\downarrow}$ inherits the phase $-\phi_{n}$ which, from eqn.~(\ref{grd}),
causes the ground state to develop the Berry phase $\Gamma=-\sum_{n}
\phi_{n}$. Because the electrons are electrically charged, one must use
the gauge-invariant form of the Berry phase \cite{ahar}
\begin{displaymath}
\phi_{n}(t)=\int^{t}_{0}\, d\tau \langle\, E_{n}|\,i\frac{d}{d\tau}+
  \frac{e}{\hbar}A_{0}(\tau)\,|E_{n}\,\rangle\hspace{0.1in}.
\end{displaymath}
$\phi_{n}(t)$ is calculated using the solutions of Ref.~\onlinecite{car},
from which one can then obtain the ground state Berry phase $\Gamma$. One
finds
\begin{equation}
\Gamma = \int\, d\tau d^{2}x\,\rho_{s}\left(\frac{1}{2}\dot{{\bf r}_{0}}
          \cdot\nabla_{{\bf r}_{0}}\theta -\frac{e}{\hbar}A_{0}\right)
           \hspace{0.1in}, \label{gbp}
\end{equation}
where ${\bf r}_{0}$ is the vortex trajectory, and we work per unit length
of the vortex. We see that our result reproduces the Berry phase obtained
in Ref.~\onlinecite{ath} for a neutral superfluid in the limit where
$e\rightarrow 0$.

We now show how the ground state Berry phase $\Gamma$ enters into the
action describing the hydrodynamic degrees of freedom of the condensate.
We begin with the vacuum-to-vacuum transition amplitude for the system
of electrons which can be written as a path integral quadratic in the
fermion fields via a Hubbard-Stratonovitch transformation
\begin{displaymath}
W=\int\, {\cal D}[\Delta]\,{\cal D}[\Delta^{\ast}]\,\langle\, vac;\Delta
   (t=T)|U_{\delta}(T,0)|vac;\Delta(0)\,\rangle \hspace{0.1in}.
\end{displaymath}
Here $U_{\Delta}(T,0)={\cal T}(\exp[-i\int_{0}^{T}\, d\tau H_{eff}])$;
$H_{eff}=H_{f}+L_{em}+L_{c}$; $H_{f}$ is the usual BCS Hamiltonian in
the presence of a 4-potential ($A_{0}$, ${\bf A}$); $L_{em}$ is the
Lagrangian for the induced electric and magnetic fields (${\bf E}$,
${\bf H}-{\bf H}_{ext}$); and $L_{c}$ is the condensation Lagrangian with
density $|\Delta|^{2}/2g$. The action for the condensate $S=S_{0}+S_{hyd}$
is given by
\begin{equation}
e^{-i(S_{0}+S_{hyd})}=\langle\, vac;\Delta(T)|U_{\Delta}(T,0)|vac; \Delta
 (0)\,\rangle \hspace{0.1in} . \label{mtx}
\end{equation}
$S_{0}$ is the action for the bulk degrees of freedom of the condensate;
$S_{hyd}$ is the action for the hydrodynamic degrees of freedom; and terms
in $S$ containing derivatives of the gap function higher than second order
are suppressed. By factoring $U_{\Delta}(T,0)$ in eqn.~(\ref{mtx}) into
a sequence of infinitesimal propogations, and appropriately inserting
complete sets of instantaneous energy eigenkets
$\{ |E_{n}(t_{k})\,\rangle\}$, evaluation of the matrix element in
eqn.~(\ref{mtx}) boils down to consideration of propogation over an
infinitesimal time interval. Spatial translational invariance, which follows
from the assumed absence of pinning sites, insures that $|vac; \Delta(0)\,
\rangle$ evolves into the instantaneous ground state $|BCS(t)\,\rangle$ of
$H_{eff}(t)$, so that the relevant matrix element is
$\langle\, BCS(t+\epsilon )|U_{\Delta(t)}(t+\epsilon ,t)|BCS(t)\,\rangle$.
One finds \cite{me}
\begin{equation}
\langle\, BCS(t+\epsilon )|U_{\Delta(t)}(t+\epsilon ,t)|BCS(t)\,\rangle =
  e^{i\Gamma\epsilon}\langle\, BCS(t)|e^{-iH_{eff}(t)\epsilon}|BCS(t)
  \,\rangle \hspace{0.1in}, \label{info}
\end{equation}
where $\Gamma$ is the Berry phase developed in $|BCS(t)\,\rangle$ due
to the vortex motion. The remaining matrix element on the RHS of
eqn.~(\ref{info}) can be evaluated \cite{eck}; and the contribution from all
infinitesimal time intervals summed. This yields the following result
for the hydrodynamic action
\begin{displaymath}
S_{hyd}=\int\, d\tau\,\left[ -\hbar\Gamma + \int\, d^{2}x\,\left[\,
         \frac{m\rho_{s}}{2}{\bf v}_{s}^{2}+N(0)\tilde{A_{0}}^{2}+
          \frac{1}{8\pi}\left\{\,\left({\bf H}-{\bf H}_{ext}\right)^{2}
           -{\bf E}^{2}\,\right\}\,\right]\,\right] \hspace{0.1in},
\end{displaymath}
in which the ground state Berry phase $\Gamma$ appears as a consequence
of the adiabatic motion of the vortex. Here ${\bf v}_{s}=-(\hbar/2m)[\nabla
\phi + (2e{\bf A})/(\hbar c)]$; $\phi$ is the gap phase; $N(0)$ is the
electron density of states at the Fermi level; $\tilde{A_{0}}=eA_{0}+
(\hbar/2)\partial_{t}\phi$; and $\hbar$, $m$ and $c$ have been re-instated.
Appropriate to the scenario of an external current passing through a thin
superconducting film in the flux-flow regime, we assume the superflow
is a combination of an applied superflow ${\bf v}= (\hbar/2m)\nabla\beta$
and one that circulates about the moving vortex with velocity
${\bf v}_{circ}=-(\hbar/2m)\nabla\theta$. The terms in $S_{hyd}$ linear
in $\nabla_{{\bf r}_{0}}\theta$ describe the coupling of the vortex to
the applied superflow ${\bf v}$; to the electric and magnetic fields via
($A_{0}$, ${\bf A}$); and to the superconducting electrons via the Berry
phase $\Gamma$. Variation of the coupling terms with respect to ${\bf r}_{0}$
gives the non-dissipative force
\begin{displaymath}
{\bf F}_{nd}=\frac{\rho_{s}h\omega}{2}\left({\bf v}-\dot{{\bf r}_{0}}\right)
              \times\hat{{\bf z}} + {\cal O}\left(\xi_{0}^{2}/\lambda^{2}
                \right) \hspace{0.1in} ,
\end{displaymath}
where $\xi_{0}$ is the zero temperature coherence length, and $\lambda$
is the London penetration depth. Our result for ${\bf F}_{nd}$ is identical
to the result found by Ao and Thouless \cite{ath} in the case of a neutral
superfluid, and which they argued would also be true for a charged
superconductor. In this first calculation we have considered the case of a
{\em charged\/} superconductor explicitly (within the context of BCS
superconductivity) and found that the Berry phase generated in the BCS
ground state {\em is\/} responsible for producing the Magnus force
contribution to ${\bf F}_{nd}$ as argued by Ao and Thouless \cite{ath}, and
that ${\bf F}_{nd}$ is given by the NV--result.
We go on now to the second {\em independent\/} calculation of ${\bf F}_{nd}$.

Our starting point (again) is the Bogoliubov equation for the case where
a line vortex with winding number $\omega = -1$ is present. We transform
the Bogoliubov Hamiltonian using the unitary operator $U=\exp[i\theta
\sigma_{3}/2]$ to obtain $H_{Bog}=\sigma_{3}[\,(i\nabla - \sigma_{3}
{\bf v}_{s})^{2}/(2) - E_{f}\,] +\Delta_{0}\sigma_{1}$. Here the $\{
\sigma_{i} \}$ are the $2\times 2$ Pauli matrices; $E_{f}$ is the Fermi
energy; and ${\bf v}_{s}=-(1/2)\nabla\theta-e{\bf A}$ ($\hbar = m=c=1$).
We make an eikonal approximation \cite{eik} for the Bogoliubov
equation eigenstates
$\phi = \exp[i{\bf q}\cdot {\bf r}]\phi^{\prime}$, where $|{\bf q}|=k_{f}$
and $\phi^{\prime}$ varies on a length scale $L\gg k_{f}^{-1}$. To first
order in gradients, this gives $H_{Bog}=\sigma_{3}[\,-{\bf q}\cdot(i\nabla
-\sigma_{3}{\bf v}_{s})]+\Delta_{0}\sigma_{1}$, from which we obtain the
gauge-invariant second quantized Lagrangian
\begin{displaymath}
{\cal L}(\hat{{\bf q}})=\Psi^{\dagger}\left[\, i\partial_{t} +\sigma_{3}
   \left(\, \frac{1}{2}\partial_{t}-eA_{0}\,\right)+\sigma_{3}{\bf q}\cdot
    \left(i\nabla - \sigma_{3}{\bf v}_{s}\right)-\Delta_{0}\sigma_{1}\,
     \right]\Psi \hspace{0.1in}.
\end{displaymath}
We see that the eikonal approximation made for the eigenstates of $H_{Bog}$
near the Fermi surface in terms of wavepackets with mean momentum
$p_{f}\hat{{\bf q}}$ has led to the separation of the $3+1$ dimensional
NQP dynamics into a collection of independent $1+1$ dimensional subsystems
labeled by directions along the Fermi surface $\hat{{\bf q}}$ and
which we will refer to as $\hat{{\bf q}}$-channels. By construction, both
positive and negative energy eigenstates (viz.\ above and below the Fermi
surface) carry a mean momentum $p_{f}\hat{{\bf q}}$. Positive energy
quasiparticles in this channel carry (mean) momentum $p_{f}\hat{{\bf q}}$
(right-goers, $\psi^{\dagger}_{R}$), while positive energy quasiholes
have (mean) momentum $-p_{f}\hat{{\bf q}}$ (left go-ers, $\psi_{L}$),
and spin indices have been suppressed. The adjoint of the NQP field
operator in this channel is $\Psi^{\dagger}_{\hat{{\bf q}}}({\bf x})
=(\psi^{\dagger}_{R}({\bf x};\hat{{\bf q}}) \;\psi_{L}({\bf x};
\hat{{\bf q}}))$. The Noether current associated with the global phase
transformation $\Psi_{\hat{{\bf q}}}\rightarrow \exp[-i\chi]
\Psi_{\hat{{\bf q}}}$ can be written in a psuedo-relativistic notation
as $j^{\mu}=\bar{\Psi}\gamma^{\mu}\Psi$. Here $\mu =0,1$; $x^{0}\equiv t$,
$x^{1}\equiv {\bf q}\cdot {\bf x}$; $\gamma^{0}\equiv \sigma_{1}$, $\gamma^{1}
\equiv -i\sigma_{2}$; and $\bar{\Psi}\equiv \Psi^{\dagger}\gamma^{0}$.
One can then write the density of linear momentum (in the
$\hat{{\bf q}}$-channel) as ${\bf g}_{i}(x;\hat{{\bf q}})=
p_{f}\hat{{\bf q}}_{i}j^{0}(x)$; and the associated stress tensor as
$T_{ij}(x;\hat{{\bf q}})=p_{f}\hat{{\bf q}}_{i}\hat{{\bf q}}_{j}j^{1}(x)$.
Taking the expectation value of these operators with respect to the
$\hat{{\bf q}}$-channel ground state $|vac\,\rangle_{\hat{{\bf q}}}$, and
summing over all $\hat{{\bf q}}$-channels gives the ground state density
of linear momentum ${\bf g}_{i}({\bf x})$ in the condensate and its
associated stress tensor $T_{ij}({\bf x})$
\begin{equation}
{\bf g}_{i}({\bf x})  =  k_{f}^{3}\sum_{\alpha}\,\int\,\frac{d\hat{{\bf q}}}
                            {4\pi^{2}}\,\hat{{\bf q}}_{i}\,\langle\,
                             vac|j^{0}(x;\hat{{\bf q}})|vac\,
                  \rangle_{\hat{{\bf q}}} \hspace{0.4in} ; \hspace{0.4in}
T_{ij}({\bf x})  =  k_{f}^{3}\sum_{\alpha}\,\int\,\frac{d\hat{{\bf q}}}
                       {4\pi^{2}}\,\hat{{\bf q}}_{i}\hat{{\bf q}}_{j}\,
                        \langle\, vac|j^{1}(x;\hat{{\bf q}})|vac \,
                         \rangle_{\hat{{\bf q}}} \label{defs}\hspace{0.1in} ;
\end{equation}
where $\alpha$ is the spin index ($\pm$). The continuity equation for the
condensate linear momentum is then
\begin{equation}
\partial_{t}g_{i} + \partial_{j}T_{ij}=k_{f}^{3}\sum_{\alpha}\,\int\,
   \frac{d\hat{{\bf q}}}{4\pi^{2}}\,\hat{{\bf q}}_{i}\,\langle\, vac|
    \partial_{\mu}j^{\mu}|vac \,\rangle_{\hat{{\bf q}}}
     \label{conty} \hspace{0.1in}.
\end{equation}
The matrix element appearing in eqn.~(\ref{conty}) does not vanish, signaling
that the condensate linear momentum is not conserved (not surprising
since the condensate is not isolated). We will see shortly that, together with
the expected source terms due to gradients in the chemical potential, and from
the electric and magnetic fields; there will also be a source term whose
origin lies in the vortex topology and which keeps track of the rate at
which linear momentum is disappearing into the vortex. This topological term
will thus give ${\bf F}_{nd}$ in this second approach. Details of the
calculation of $M=\langle\, vac|\partial_{\mu}j^{\mu}|vac\,
\rangle_{\hat{{\bf q}}}$ are given in Ref.~\onlinecite{me}. The result is
$M=(\epsilon^{\mu\nu}\tilde{F}_{\mu\nu})/4\pi$, where $\epsilon^{01}=1$;
$\tilde{F}_{\mu\nu}=\partial_{\mu}\tilde{A}_{\nu}-\partial_{\nu}
\tilde{A}_{\mu}$; and $\tilde{A}_{0}=eA_{0}-(1/2)\partial_{t}\theta$;
$\tilde{A}_{1}={\bf q}\cdot {\bf v}_{s}$. Inserting this result for $M$ into
eqn.~(\ref{conty}) gives
\begin{equation}
\partial_{t}g_{i}+\partial_{j}T_{ij}= C_{0}\left\{\, -\frac{\hbar}{2}
  \left[\,\partial_{0}, \partial_{i}\,\right]\theta +e{\bf E}_{i}\,\right\}
   \hspace{0.1in}. \label{newc}
\end{equation}
Here $C_{0}=k_{f}^{3}/3\pi^{2}$ is the particle density in the normal phase
for the case where the chemical potential equals the Fermi energy; and $\hbar$
has been re-stored. The first term on the RHS of eqn.~(\ref{newc}) is
non-vanishing due to the non-trivial vortex topology. The local expression
of this topology, appropriate for a vortex with winding number $\omega$, is
$[\partial_{x},
\partial_{y}]\phi = 2\pi\omega\delta (x-x_{0})\delta (y-y_{0})$, where
$\phi$ is the gap phase and ${\bf r}_{0}= (x_{0},\; y_{0})$ is the position
of the vortex.
The LHS of eqn.~(\ref{newc}) can also be evaluated using eqn.~(\ref{defs})
along with the result \cite{me} $\langle\, vac|j^{\mu}|vac \,
\rangle_{\hat{{\bf q}}}=\epsilon^{\mu\nu}\tilde{A}_{\nu}/2\pi$. The results
are
$g_{i}=C_{0}\left({\bf v}_{s}\right)_{i}$;
$T_{ij}=C_{0}\left(\frac{\hbar}{2}\partial_{t}\theta -
eA_{0}\right)\delta_{ij}$.
Making use of these results in eqn.~(\ref{newc}); together with the Josephson
equation $(\hbar\partial_{t}\phi)/2 = -\mu_{0}$, where $\mu_{0}$ is the
chemical
potential in the vortex rest frame which can be written as $\mu_{0}=
\mu + {\bf v}_{s}^{2}/2 +eA_{0}$ ($\mu$ is the chemical potential in the
lattice frame and $m=1$) gives finally
\begin{equation}
\frac{d{\bf v}_{s}}{dt}=-\nabla\mu +e{\bf E}+e{\bf v}_{s}\times {\bf B}
  -\frac{h\omega}{2}\left({\bf v}_{s}-\dot{{\bf r}_{0}}\right)\times
    \hat{{\bf z}}\,\delta^{2}({\bf r}-{\bf r}_{0}) \label{acc}\hspace{0.1in}.
\end{equation}
We see that the continuity equation for the condensate linear momentum has
yielded the acceleration equation for the superflow. We find the expected
source terms related to the hydrodynamic pressure ($\nabla P=\rho_{s}\nabla
\mu$), and the electric and magnetic fields. We also see that linear
momentum is disappearing from the condensate into the vortex at ${\bf r}_{0}
(t)$ at the rate $(\rho_{s}h\omega /2)({\bf v}_{s}-\dot{{\bf r}_{0}})\times
\hat{{\bf z}}$ per unit length so that
\begin{displaymath}
{\bf F}_{nd}=\frac{\rho_{s}h\omega}{2}\left({\bf v}_{s}-\dot{{\bf r}_{0}}
  \right)\times\hat{{\bf z}}\hspace{0.1in},
\end{displaymath}
in agreement with the Berry phase calculation. Our result is
also consistent
with the calculation of NV in Ref.~\onlinecite{phen}. These authors showed
that the first 3 terms in eqn.~(\ref{acc}) lead to a flux of linear momentum
in towards the vortex at a rate $(\rho_{s}h\omega )/(2)({\bf v}_{s}-
\dot{{\bf r}_{0}})\times\hat{{\bf z}}$ which is exactly the rate at which
we find it appearing on the vortex, indicating that linear momentum is
conserved in the combined condensate-vortex system.

In this paper we have provided two {\em independent\/} microscopic
calculations
of the non-dissipative force ${\bf F}_{nd}$ acting on a line vortex in a
type-II superconductor at $T=0$. Both calculations yield the
NV-form for this force ${\bf F}_{nd}=(\rho_{s}h\omega /2)
({\bf v}-\dot{{\bf r}_{0}})\times\hat{{\bf z}}$. The first calculation
(inspired by earlier work of Ao and Thouless which determined
${\bf F}_{nd}$ via a Berry phase analysis appropriate for a neutral
superfluid, and which they argued would also be true for a charged
superconductor) shows that the arguments of Ao and Thouless are fully
borne-out in the context of the BCS model for a {\em charged\/}
superconductor. The second calculation (which does not
rely on Berry phases) examines the flow of linear momentum in the condensate.
The continuity equation for this linear momentum is shown to: (i) yield the
acceleration equation for the superflow; and (ii) to contain a sink term
indicating the disappearance of linear momentum into the vortex. ${\bf F}_{nd}$
follows in this second approach from the rate of momentum loss to the vortex.
The result obtained is the NV result and the Magnus force contribution to
${\bf F}_{nd}$ is seen to be a consequence of the vortex topology.

{\em Note Added\/}: Two preprints have appeared since this work was completed
(M. Stone; Aitchison et.\ al.\ \cite{prprt}) which also find a
gauge-invariant contribution to the hydrodynamic action first order in
time derivatives of the gap phase.

I would like to thank Ping Ao for interesting me in this problem and for
helpful discussions; Michael Stone for interesting comments and discussions,
particularly with regard to the issue of gauge invariance; T. Howell III
for constant support; and NSERC of Canada for financial support.


\begin{references}
\bibitem{defs} We assume a line vortex with axis along $\hat{{\bf z}}$;
               $\rho_{s}$ is the superconducting electron density at
               $T=0$ far from the vortex; $h$ is Planck's constant;
               $\omega = \pm 1$ is the vortex winding number and
               corresponds to a single flux quantum $\phi_{0}=hc/2e$
               threading the vortex; ${\bf v}_{s}$ is the velocity of
               an applied supercurrent with respect to the lattice rest
               frame; $m$ is the electron mass; ${\cal K}=h\omega /2m$
               is the
               circulation of the condensate near the vortex core; and
               ${\bf v}_{L}$ is the translational velocity of the vortex
               with respect to the lattice.
\end{references}
\end{document}